\documentclass[preprint,nofootinbib,amsmath,amssymb,floatfix,showpacs]{revtex4}
\usepackage{graphicx}
\begin{document}

\title{Comment on the evidence of the Higgs boson at LHC} 

\author{Paolo Cea$^{1,2}$}
\email[]{Paolo.Cea@ba.infn.it}

\affiliation{$^1$Dipartimento di Fisica dell'Universit\`a di Bari, I-70126 Bari, Italy \\
$^2$INFN - Sezione di Bari, I-70126 Bari, Italy }

\begin{abstract}
We comment on the Standard Model Higgs boson evidence from LHC.  We propose that the new resonance at 125 GeV could be
interpreted as a pseudoscalar  meson with quantum number $J^{PC} = 0^{- +}$. We show that this pseudoscalar 
could mimic the decays of the Standard Model Higgs boson in all channels with the exception of the decay into
two leptons that is strongly suppressed due to charge-conjugation invariance. 
\end{abstract}


\maketitle

\begin{figure}[t]
%
%
\includegraphics[width=1.0\textwidth,clip]{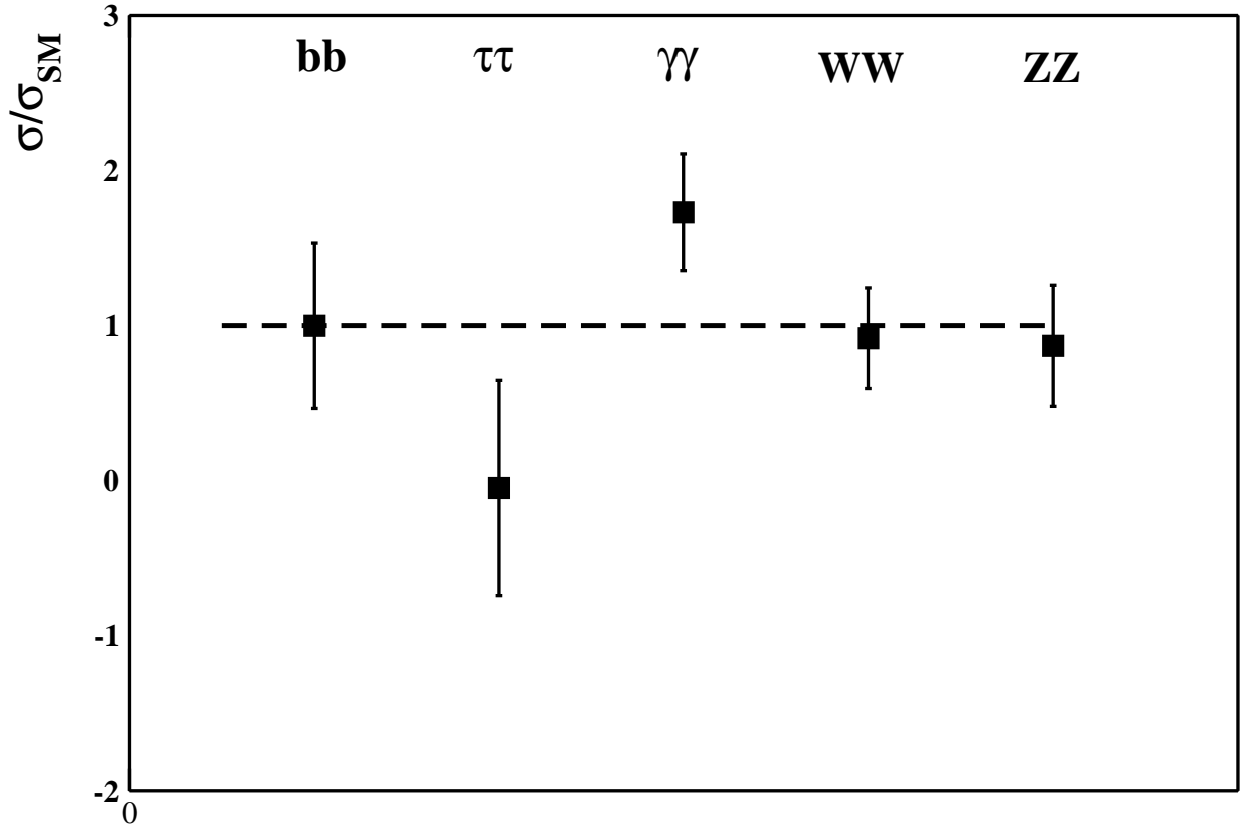}
\caption{\label{Fig-1}  Values of signal strength  for individual decay modes. The ratio $\frac{\sigma}{\sigma_{\text{SM}}}$ denotes
the production cross section times the relevant branching fractions  relative to the Standard Model expectations.
The data are based on the combination of the  LHC~\cite{Aad:2012,Chatrchyan:2012} and  TEVATRON~\cite{Shalhout:2012}  data.
The dashed line is the Standard Model value  $\frac{\sigma}{\sigma_{\text{SM}}} = 1$. }
%
%
\end{figure}

Recently  the ATLAS~\cite{ATLAS:2012,Aad:2012} and CMS~\cite{CMS:2012,Chatrchyan:2012} experiments at the LHC reported an evidence of a new particle consistent
with  the Standard Model Higgs boson~\cite{Englert:1964,Higgs:1964,Guralnik:1964,Higgs:1966} at a mass of $M_H \simeq 125$ GeV. In fact, the
evidence of a new resonance was already present in the 2011 data and has been consistently confirmed in several decay channels by both the ATLAS and CMS
experiments. \\
In order to check that the new resonance is, indeed, the Standard Model Higgs boson one may measure the rate for production of the
Higgs boson in a given decay channel by introducing the relative signal strength $\mu$, defined by:
\begin{equation}
\label{1}
\mu \; = \; \frac{\text{signal rate from fit to data}}{ \text{expected SM signal rate at given M}_H }  \;  \; ,
\end{equation}
where SM denotes the Standard Model prediction.  \\
The ATLAS~\cite{Aad:2012}, CMS~\cite{Chatrchyan:2012}, and TEVATRON~\cite{Shalhout:2012} experiments have presented values of
the signal strength   $\mu$ for various decay channels. 
In Fig.~\ref{Fig-1} we summarize the combined analyses  of ATLAS and CMS results at 7 and 8 TeV  with the Tevatron experiments data.
It is worthwhile to stress that our combination must be intended as merely indicative, for the full combination of different
experimental data is much more involved. Nevertheless, it is  amusing that the emerging picture is quite consistent with various
experimental analyses. In fact,  the signal strength data roughly resemble what would be expected of a Standard Model Higgs boson
 despite the moderate  enhancement in the diphoton channel and the deficit in $\tau^+\tau^-$ decay channel.
We see, then, that the newly discovered particle is not inconsistent with
the SM Higgs boson,  but it also may be different from the SM
Higgs boson. In fact, the true identity of any putative Higgs boson will remain ambiguous
until one has experimentally excluded other possible assignments of quantum numbers and couplings.
In particular,  we believe that the lack of signal in the leptonic decay could require an alternative explanation for
the new particle detected at the LHC. 
\\
The aim of the present note is to discuss  a possible alternative to the  generally assumed Higgs boson
interpretation. Since up to now there is  no signs of new physics, we looked for alternative explanations
within the Standard Model physics.  The absence of signal with respect to  SM background  in the
leptonic decay channels suggests that the new resonance could be
interpreted as a pseudoscalar  meson with quantum number $J^{PC} = 0^{- +}$. Indeed, for a  pseudoscalar meson  the decay into
two leptons  is naturally  suppressed due to charge-conjugation invariance. The most natural pseudoscalar candidate within the Standard Model
is a $q \bar{q}$ bound state with L=S=0. Given the large mass of the new resonance we consider the pseudoscalar $t \bar{t}$ which, for obviously reasons, 
we will refer to as $\eta_{t}$.  Since the top quark mass is very large:
\begin{equation}
\label{2}
m_t \;  \simeq  \;  173 \; \text{GeV} \; ,
\end{equation}
to estimate the mass of the pseudoscalar meson  $\eta_{t}$, we may safely employ the non-relativistic potential model.
Quarkonium potential models typically take the form of a Schr\"odinger like equation:
\begin{equation}
\label{3}
[T \; + \;  V ] \; \Psi	\;  =   \;  m \; \Psi
\end{equation}
where T represents the kinetic energy term and V the potential energy term.
The $\text{quark - antiquark}$  potential is typically motivated by the properties expected from QCD.
At short distances one-gluon-exchange leads to the Coulomb
like potential. At large distances the one-gluon-exchange is no longer a good representation of the $\text{quark - antiquark}$  potential. The qualitative
picture is that the chromoelectric lines of force bunch together into a flux tube which leads to a distance independent
force or linearly rising confining potential. To be definite we will use
the  so-called Cornell potential~\cite{Eichten:1980}:
\begin{equation}
\label{4}
V_C  = -\; \frac{4}{3} \, \frac{\alpha_s }{r} \; + \; \sigma \, r  \; , \; \alpha_s \, \simeq \, 0.40 \; , \; \sigma  \simeq \, 0.18 \; \text{GeV}^2  .
\end{equation}
We are interested in a qualitative estimate of the low-lying L=S=0 bound state. Since the contribution of the linearly rising confining potential
can be safely neglected due to the very large top mass, we  obtain at once  the wave function of the low-lying L=S=0 bound state:
\begin{equation}
\label{5}
\Psi_{00}(r) \; \simeq \;  \frac{1}{ ( \pi \;  a_0^3)^{\frac{1}{2}}} \;  \exp(-\frac{r}{a_0} )  \; \; ,
\end{equation}
where $a_0$ is the Bohr radius:
\begin{equation}
\label{6}
a_0 \; =  \;  \frac{ 3}{2 \; m_t \; \alpha_s}    \; \; .
\end{equation}
We may, then, estimate the pseudoscalar mass as follows:
\begin{equation}
\label{7}
m_{\eta_{t}}  \; \simeq  \; 2 \; m_t \; -\; \frac{4}{3} \; \frac{\alpha_s }{a_0} \;  \simeq  \;  321 \; \text{GeV}  \; .
\end{equation}
Even though our analysis has been somewhat qualitative, it is evident that the pseudoscalar $\eta_{t}$ meson is too heavy to be identified with the
new LHC resonance. To overcome this problem we must admit that the $\eta_{t}$ meson can have sizable mixing with a much more  lighter 
 pseudoscalar meson. During the completion of the present note, we were aware of Ref.~\cite{Moffat:2012} were it is suggested that
the new resonance could originate from a sizable mixing between the pseudoscalar $t \bar{t}$ meson ($\eta_{t}$) and the pseudoscalar $b \bar{b}$
meson ($\eta_{b}$).  It should be stressed that for  $q \bar{q}$  pseudoscalar states it is believed~\cite{DeRujula:1975} that the mixing is expected to proceed via
the annihilation into two gluons which form a color singlet. However, this annihilation amplitude is strongly suppressed
by the large top mass.
In fact, the experimental observation  of the pseudoscalar $\eta_{b}$ meson at the mass  $m_{\eta_{b}} = 9391.0 \pm 2.89 \;  \text{MeV}$~\cite{Beringer:2012} implies that the $\eta_{b} - \eta_{t}$ mixing (if any) is completely negligible. On the other hand, the self-coupling of gluons in QCD suggests that additional
mesons made of bound gluons (glueballs) may exist. In fact, lattice calculations, flux tube and constituent glue models agree that the lights glueballs 
have quantum number $J^{PC} = 0^{++}, 2^{++}$  (for a recent review see Refs.~\cite{Klempt:2007,Crede:2009}). Moreover, there is a general
agreements on the existence of pseudoscalar states with $J^{PC} = 0^{-+}$  above 2 GeV. In the following we will indicate the lowest glueball
pseudoscalar state with $\eta_{g}$ and follow the lattice calculations for the mass of the lowest pseudoscalar glueball  to set the value~\cite{Crede:2009}: 
\begin{equation}
\label{8}
m_{\eta_{g}} \;  \simeq  \;  2.6 \; \text{GeV} \; .
\end{equation}
We see, then, that the pseudoscalar $\eta_{t}$ meson can also mix with the pseudoscalar  $\eta_{g}$ meson through color singlet gluon intermediate
states. In this case there are no reasons to restrict the intermediate states to two gluons. In fact, if more gluons are involved then one gets large
effective couplings due to the small typical momentum going into each one making the theory strongly coupled~\cite{Appelquist:1975}. If this
is the case, then the large top mass gives rise to a sizable mixing amplitude. 
We shall proceed as in Ref.~\cite{DeRujula:1975} for the  mesons $\eta$ and $\eta'$. In fact,  if we assume that the annihilation process contribute
the flavor independent amount A, we obtain the following mass matrix:
\begin{equation}
\label{9}
\cal{M} \; = \;  \left(
\begin{array} {cc}
m_{\eta_g} \; + \; A & A \\
A & m_{\eta_t}  \; + \; A
\end{array}
\right) \; \;.
\end{equation}
The mass matrix can be easily diagonalized by writing the physical mass eigenstates as:
\begin{eqnarray}
\label{10}
 \eta_{gt} \;  & = &    \eta_g \; cos \theta  \; \; - \; \; \eta_t  \; sin \theta  
 \nonumber   \\
  \\
 \eta'_{gt} \;  & = &    \eta_g \; sin \theta  \; \; + \; \; \eta_t  \; cos \theta  \;
 \nonumber
\end{eqnarray}
where $\theta$ is the mixing angle. Inverting Eq.~(\ref{10}) leads to:
\begin{eqnarray}
\label{11}
 \eta_{g} \;  & = &    \eta_{gt} \; cos \theta  \; \; + \; \; \eta'_{gt}  \; sin \theta  
 \nonumber   \\
  \\
 \eta_{t} \;  & = &    \eta'_{gt} \; cos \theta  \; \; - \; \; \eta_{gt}  \; sin \theta   \; \; .
 \nonumber
\end{eqnarray}
We denote with $\eta_{gt}$ the state with lowest mass eigenvalue. Moreover we impose that: 
\begin{equation}
\label{12}
m_{\eta_{gt}} \;  \simeq  \;  125 \; \text{GeV} \; .
\end{equation}
A standard calculation gives:
\begin{equation}
\label{13}
m_{\eta'_{gt}} \;  \simeq  \;  850 \; \text{GeV} \;  \; \; , \; \; \; \theta \; \simeq \; 19 \,^\circ  \; \; .
\end{equation}
Since the mass eigenstate $\eta'_{gt}$ lies well above the $t \bar{t}$ threshold, it hardly can be detected as a hadronic resonance. On the other hand,
the eigenstate $\eta_{gt}$ could be a serious candidate for the new resonance detected at LHC. To check this, we need to estimate the total width and the
decay channels of the pseudoscalar meson $\eta_{gt}$.  Obviously, the main channels  are given by the decay of  $\eta_{gt}$ into ordinary hadrons, which
are suppressed by the OZI rule. To estimate the contribution of the   $\eta_{t}$  component to the  decay width we may use well known heavy quarkonium
model~\cite{Close:1979}:
\begin{equation}
\label{14}
\Gamma[\eta_{t} \rightarrow hadrons ]  \simeq  \Gamma[\eta_{t} \rightarrow g g ]  \simeq  
\frac{32 \pi}{3} \frac{\alpha_s^2(m_{\eta_{t}})}{m_{\eta_{t}}^2} | \Psi_{00}(0)|^2 ,
\end{equation}
where $\Psi_{00}$ is given by Eq.~(\ref{5}). Note that the decay  into hadrons is proportional
to $\alpha_s^2$ since by  charge conjugation conservation $\eta_t$ can decay into two gluons. 
Since the hadronic decay width depends on the wave function, we see that the glueball contributions
can be safely neglected. This is consistent with the general expectation that glueball should  be narrower that 
$q \bar{q}$ mesons. Thus we obtain:
\begin{equation}
\label{15}
\Gamma[\eta_{gt} \rightarrow  \eta_t \rightarrow hadrons ]  \; \simeq  \; \Gamma[\eta_{t} \rightarrow g g ]  \; \sin^2 \theta  \; .
\end{equation}
As concern the coupling of the pseudoscalar meson $\eta_{gt}$ to the  gauge vector bosons, we note that
 the  glueballs are made of electrically neutral gluons, so that naively we expect suppressed couplings to the gauge vector bosons.
Thus  we may  neglect the glueball contributions to the $\eta_{gt}$ decay.  We shall return on this subject later on.
The most sizable decay should be the decay into two photons. Accordingly we get:
\begin{equation}
\label{16}
\Gamma[\eta_{gt} \rightarrow  \eta_t \rightarrow \gamma \gamma]  \; \simeq  \; \Gamma[\eta_{t} \rightarrow \gamma \gamma]    \; \sin^2 \theta  \; ,
\end{equation}
where ~\cite{Close:1979}:
\begin{equation}
\label{17}
\Gamma[\eta_{t} \rightarrow \gamma \gamma]  \simeq 
\frac{256 \, \pi}{27} \;  \frac{\alpha^2(m_{\eta_{t}})}{m_{\eta_{t}}^2}  \; | \Psi_{00}(0)|^2  \; .
\end{equation}
Assuming:
\begin{equation}
\label{18}
  \alpha_s(m_{\eta_{t}}) \; \simeq \; 0.10 \; \; \; , \; \; \;     \alpha(m_{\eta_{t}})   \simeq   \; \frac{1}{127} \; ,
\end{equation}
we get:
\begin{equation}
\label{19}
\Gamma[\eta_{gt} \rightarrow  \eta_t  \rightarrow hadrons ] \;  \simeq  \; 11  \; \text{MeV}  \;  \; , 
\end{equation}
\begin{equation}
\label{20}
\Gamma[\eta_{gt}  \rightarrow  \eta_t \rightarrow \gamma \gamma ]   \; \simeq  \;  60  \; \text{KeV}  \; \;  .
\end{equation}
Up to now we have neglected the glueball component of the pseudoscalar meson $\eta_{gt}$ on the 
basis that glueballs should  be narrower that  ordinary $q \bar{q}$ mesons. This general rule
has an exception for the $\gamma \gamma$ coupling. In fact, it is known since long time that
glueballs may have $\gamma \gamma$ couplings as large as $q \bar{q}$ mesons~\cite{Chanowitz:1984}.
This remarkable property follows from the trace and chiral anomalies.
Indeed, from  low energy theorems one obtains  for the
$\gamma \gamma$ widths of the lightest pseudoscalar glueball~\cite{Chanowitz:1984}:
\begin{equation}
\label{21}
\Gamma[\eta_{g} \rightarrow \gamma \gamma]  \simeq 
\frac{\alpha^2}{24 \, \pi^3} \;  \frac{m_{\eta_g}^3}{f_{\eta_g}^2}  \;  \; ,
\end{equation}
where $f_{\eta_g}$  is the analogous of the pion decay constant $f_\pi$~\cite{Scadron:1981}. 
Accordingly, we may write:
\begin{equation}
\label{22}
\Gamma[\eta_{gt} \rightarrow  \eta_g \rightarrow \gamma \gamma]  \; \simeq  \; \Gamma[\eta_{g} \rightarrow \gamma \gamma]    \; \cos^2 \theta  \; .
\end{equation}
To estimate the contribution to the $\gamma \gamma$ decay width of the glueball component of the pseudoscalar meson $\eta_{gt}$,
we assume  $f_{\eta_g} \simeq f_\pi$. Thus, from Eqs.~(\ref{21}) and (\ref{22}) we get:
\begin{equation}
\label{23}
\Gamma[\eta_{gt} \rightarrow \gamma \gamma ]   \; \simeq  \;  110  \; \text{KeV}  \; \;  .
\end{equation}
Equation~(\ref{23}) confirms that, indeed, the the glueball component of the pseudoscalar meson $\eta_{gt}$
gives a contribute to  the $\gamma \gamma$ decay width of the same order as the $t\bar{t}$ component, Eq.~(\ref{20}). \\
To summarize, the main decay channels of our pseudoscalar meson are determined by the heavy quarkonium component
with the exception of the $\gamma \gamma$ decay where there is a sizable contribution from the glueball component.
Note that this jeopardizes the recent phenomenological analysis presented in Ref.~\cite{Coleppa:2012}. 
We see, then, that this peculiar pseudoscalar 
could mimic the decays of the Standard Model Higgs boson in all channels with the exception of the decay into
two leptons that turns out to be naturally suppressed. 
Indeed, the decays of  the pseudoscalar meson $\eta_{gt}$  into charged leptons proceeds
mainly through  two virtual photons  by  charge conjugation conservation. Thus, as an order of magnitude  estimate, we obtain:
\begin{equation}
\label{29}
\frac{\Gamma[\eta_{gt} \rightarrow \ell^+ \ell^- ] }{\Gamma[\eta_{gt} \rightarrow \gamma \gamma ] } \;  \sim \;
 \alpha^2 \; \sim \; 10^{-4}  \; .
\end{equation}
In conclusion we are suggesting that to identify the new LHC resonance with the Standard Model Higgs boson
it is of fundamental importance to determine experimentally both the spin and the parity. Fortunately,
angular analysis based with analytical likelihood can separate pseudoscalar-scalar hypotheses at a $ 3 \sigma$  level 
with  about 30 $\text{fb}^{-1}$~\cite{Peskin:2012}. So that the forthcoming data from LHC will be enough to
fix the  spin and parity of the new resonance. \\
If the new state at 125 GeV should turn out to be a pseudoscalar, then one faces with the problem of the spontaneous symmetry
breaking mechanism and the related scalar Higgs boson. 
In a recent paper~\cite{Cea:2012} the scenario   is discussed  where the Higgs boson without self-interaction (Trivial Higgs) could coexist with spontaneous symmetry breaking.  Due to the peculiar rescaling of  the Higgs condensate, the relation between Higgs mass $m_H$ and the physical Higgs condensate  $v_R$ 
is not the same as in perturbation theory.  
According to this picture   one expects that  the ratio $m_H/v_R$ would  be a cutoff-independent constant.  In fact, the lattice studies~\cite{Cea:2012}
 showed that the extrapolation to the continuum limit leads to the quite simple result:
\begin{equation}
\label{30}
m_H \; \simeq \;  \pi  \;  v_R  
\end{equation}
pointing to  a rather massive Trivial Higgs boson $m_H  \simeq 750$  GeV. 

\end{document}